\documentclass[11pt,a4paper]{article}

\pdfoutput=1
\usepackage{jheppub}

\usepackage{epsfig}
\usepackage{hyperref}
\usepackage{amssymb}
\usepackage{amsbsy}
\usepackage{amsmath}
\usepackage{url}

\newcommand{\rf}[1]{(\ref{#1})}
\newcommand{\beq}{\begin{equation}}
\newcommand{\eeq}{\end{equation}}
\newcommand{\be}{\begin{equation}}
\newcommand{\ee}{\end{equation}}
\newcommand{\bea}{\begin{eqnarray}}
\newcommand{\eea}{\end{eqnarray}}
\newcommand{\eq}[1]{Eq.~(\ref{#1})}
\newcommand{\non}{\nonumber \\*}
\newcommand{\ie}{{i.e.}\ }
\newcommand{\bv}{\begin{verbatim}}
\newcommand{\ev}{\end{verbatim}}

\newcommand{\vp}{\varphi}

\newcommand{\e}{\,\mbox{e}}
\renewcommand{\d}{{\rm d}}
\renewcommand{\i}{{\rm i}}
\newcommand{\blambda}{\bar\lambda}
\newcommand{\brho}{\bar\rho}

\newcommand{\bz}{{\bar z}}
\newcommand{\p}{\partial}
\newcommand{\bp}{\bar\partial}
\newcommand{\q}{\mbox{}}
\newcommand{\qq}{q}

\newcommand{\eps}{\varepsilon}
\newcommand{\om}{\omega}

\newcommand{\LA}{\left\langle}
\newcommand{\RA}{\right\rangle}

\def\fun#1#2{\lower3.6pt\vbox{\baselineskip0pt\lineskip.9pt
\ialign{$\mathsurround=0pt#1\hfil##\hfil$\crcr#2\crcr\sim\crcr}}}
\begin{document}

\preprint{}

%\title{Note on conformal theory with nonprimary energy-momentum tensor \\ that describes the Nambu-Goto  string}

\title{Notes on higher-derivative conformal theory with nonprimary energy-momentum tensor
 that applies to the Nambu-Goto  string}

\author
{Yuri Makeenko}
\vspace*{2mm}
\affiliation{NRC ``Kurchatov Institute''\/-- ITEP, Moscow\\
%\affiliation{Institute of Theoretical and Experimental Physics, Moscow\\
%%\mbox{B. Cheremushkinskaya 25, 117218 Moscow, Russia}\\
\vspace*{1mm}
{makeenko@itep.ru} 
}
%\today

\begin{abstract}
{I investigate the higher-derivative conformal theory  which 
shows how the Nambu-Goto and Polyakov strings can be told apart.
Its energy-momentum tensor is conserved, traceless 
%%but is not a primary conformal field of BPZ.
but does not belong to the conformal family of the unit operator.
To implement conformal invariance in this case, I develop the new technique that 
explicitly accounts for the quantum equation of motion and results in singular products.
I show that the conformal transformations generated by such a nonprimary energy-momentum tensor
form a Lie algebra with a  central extension 
which in the path-integral formalism gives a logarithmically  divergent contribution 
to the central charge. 
I demonstrate  how the logarithmic divergence is canceled in the string susceptibility  and reproduce the previously obtained deviation from KPZ-DDK at one loop.}

\end{abstract}

%\pacs{11.25.Pm,11.25.Hf, 11.15.Pg} 

\maketitle

 \section{Introduction}
 
I continue in these Notes the investigation \cite{Mak22,Mak21} of conformal symmetry of the Nambu-Goto string. My original motivation for this study was the celebrated formula
\be
\gamma_{\rm str}=(1-h)\left[\frac{d-25-\sqrt{(25-d)(1-d)}}{12}\right]+2
\label{ggg}
\ee
by Knizhnik-Polyakov-Zamolodchikov~\cite{KPZ}  and
David-Distler-Kawai~\cite{DDK} (abbreviated as KPZ-DDK)
for the string susceptibility index $\gamma_{\rm str}$ of surfaces of genus $h$. 
It was derived for the Polyakov string in $d$ target-space dimensions using its conformal
invariance
and suffers the so-called $d=1$ barrier above which \rf{ggg} is not real
and thus unacceptable. This is sometimes formulated as a no-go theorem for 
the existence of bosonic string at $d>1$.

The results of Ref.~\cite{Mak22} suggest that a potential way out of this problem could be 
the fact that the Nambu-Goto and the Polyakov strings are in fact not equivalent despite of 
original Polyakov's argument~\cite{Pol87}.
The origin of this non-equivalence can be seen already at the level of the simplest  
higher-derivative beyond Lioville action
\be
{\cal S}[\vp]=\frac 1{16 \pi b_0^2} \int \sqrt{g} \left[g^{ab} \p_a \vp \p_b \vp  +2m_0^2
+\eps  R \left(R+ G g^{ab}\,\partial_a \vp\partial_b \vp \right)\right] .
%%,\quad b^2_0=\frac 6{26-d},
\label{inva}
\ee
%which emerges after integrating over the target-space coordinates and ghosts.
%, with the parameter $\eps$ being proportional to the ultraviolet cutoff in the target space. 
Here $R$ is the scalar curvature for the two-dimensional metric tensor $g_{ab}$ and
$\vp=-\Delta^{-1} R$ (with $\Delta$ being the two-dimensional Laplacian)
becomes a local field in the conformal gauge 
\be
g_{ab}=  \hat g _{ab} \e^\vp ,
\label{confog}
\ee
where $\vp$ is the dynamical variable
often called the Liouville field and $\hat g_{ab}$ is a slowly varying  background (fiducial) metric tensor.  
The term $R^2$ appears already for the Polyakov  string but the second higher-derivative term 
with $G\neq 0$ is specific to the Nambu-Goto string~\cite{Mak21},
as will be reviewed in the next section.
%with the {worldsheet cutoff} $\eps=a^2/\brho$ and $\mu_0^2=m_0^2 \brho$.

When the action~\rf{inva} emerges as an effective action from a string by path-integrating over
the target-space coordinates, ghosts 
(and the Lagrange multiplier in the case of the Nambu-Goto string), the parameter $\eps$ is proportional to the ultraviolet cutoff
in the target space. Thus the higher-derivative terms in the action~\rf{inva} 
 are suppressed for
smooth metrics as $\eps R$. However, typical metrics which are essential in the
path integral over the metrics $g_{ab}$ are not smooth and have $R\sim \eps^{-1}$, 
so the higher-derivative terms revive~\cite{Mak21} after doing uncertainties like $\eps \times \eps^{-1}$. 
I shall consider in these Notes the action~\rf{inva} as such not assuming that $\eps$ is infinitesimally small.

A very interesting property of the action~\rf{inva} is that
the associated energy-momentum tensor $T_{ab}^{(\vp)}$ is conserved and traceless~\cite{Mak22}
owing to the classical equation of motion 
\be
-\Delta \vp+m_0^2 + \eps \Delta^2 \vp -\frac{\eps}2 (\Delta \vp)^2  
+\frac12 G\eps \p^a \vp \p_a \vp \Delta \vp-\frac12 G\eps \p^a\p_a(\p^b\vp\p_b\vp)+
 G\eps \p_a  (\p^a \vp \Delta \vp)=0
\label{cemG}
\ee
%\be
%-\Delta \vp+m_0^2 + \eps \Delta^2 \vp -\frac{\eps}2 (\Delta \vp)^2  
%+\frac12 G\eps g^{ab} \p_a \vp \p_b \vp \Delta \vp
%-\frac12 G\eps g^{ab}\p_a\p_b(g^{cd}\p_c\vp\p_d\vp)+
%G\eps \p_a  (g^{ab}\p_b \vp \Delta \vp)=0
%\label{cemG}
%\ee
despite  $m_0$ and $\eps$ are dimensionful. 
%This is of course a consequence of symmetries of the action \rf{inva}. 
The model  described by the action \rf{inva} thus possesses conformal symmetry at
least at the classical level.

 For this reason a
 great simplification of the  formulas occurs when using the conformal coordinates $z$ and $\bz$, 
 where the flat metric tensor becomes
\be
\hat g_{ab}=\left(
\begin{array}{cc}
0 ~&~\frac12 \\
\frac12 ~&~ 0\\
\end{array}
\right),\qquad
\hat g^{ab}=\left(
\begin{array}{cc}
0 ~&~2 \\
2 ~&~ 0\\
\end{array}
\right).
\ee
The  $T_{zz}$ and $T_{z\bz}$ components of
the energy-momentum tensor  $T_{ab}^{(\vp)}$ then read
\bea
-4 b_0^2 T_{zz}^{(\vp)}&=&(\p \vp)^2 -2\eps \p \vp \p \Delta \vp -2\q\p^2 (\vp - \eps\Delta \vp)
-G \eps (\p \vp)^2 \Delta \vp+ 4G\eps\p\vp \p(\e^{-\vp}\p \vp \bp \vp)  \non
 && 
- 4G\q\eps \p^2(\e^{-\vp}\p \vp \bp \vp)
 +  G\q \eps\partial(\p \vp \Delta\vp)+
  G \q\eps \frac 1{\bp}\p ^2 (\bp \vp \Delta\vp),\qquad \Delta=4\e^{-\vp}\p  \bp 
  \label{Tzz} \non &&
\eea
and $b_0^2 T_{z\bz}^{(\vp)}\!\e^{-\vp}$ given by the left-hand side of \eq{cemG} in the conformal gauge.
We used the notation $\partial\equiv \partial/\p z$ and
$\bar \partial\equiv \partial/\p\bz$.
Notice the nonlocality of the last term in \rf{Tzz} which is inherited from a
nonlocality of the action \rf{inva}. 
The presence of this nonlocal term plays a crucial role in the computation of 
the central charge at one loop~\cite{Mak22}.

$T_{zz}$ given by \rf{Tzz} obeys the conservation law%
\footnote{This can be verified using %%playing with 
the Mathematica program from Appendix~\ref{appA}.}
\be
\bar \partial T_{zz}^{(\vp)}=0
\label{conserv}
\ee
because  $T_{z\bz}^{(\vp)}$  %%given by \rf{Tzbz} 
vanishes owing to the classical equation of motion \rf{cemG}.  
In the quantum case \eq{cemG} is replaced by the Schwinger-Dyson equation
\be
\hbox{left-hand side of \eq{cemG}}\stackrel{{\rm w.s.}}=8\pi b_0^2 \frac {\delta}{\delta \vp},
\label{SDe}
\ee
where the equality is undersood in the weak sense, \ie under the sign of averages  
in the path-integral formalism.
Equation~\rf{cemG} is recovered in the classical limit $b_0^2\to0$.

The  infrared limit of our model is described by an effective action, governing smooth fluctuations of $\vp$,
and the effective energy-momentum tensor
\be
T_{zz}^{({\rm eff})}=\frac{1}{2b^2} \left( \qq\p^2 \vp -\frac 12 (\p\vp)^2 \right)
\label{Teff}
\ee
which is %%%also 
quadratic in $\vp$. The arguments are similar to %%David-Distler-Kawai
 DDK~\cite{DDK}. 
Here $b^2$ describes the
renormalization of $\vp$, \ie the change $b_0^2\to b^2$ in the action \rf{inva} and $q$ characterizes the theory.
In the usual case of the Liouville action where $\eps=0$ they obey the DDK equation
\be
\frac{6q^2}{b^2}+1=\frac{6}{b_0^2}
\label{DDK1}
\ee
derived from the background independence. The left-hand side of \eq{DDK1} is the central charge of $\vp$.
For the Polyakov string \eq{DDK1} provides the vanishing of the total central charge
leading for $b_0^2=6/(26-d)$ to \eq{ggg}. 

\begin{figure}
\includegraphics[width=13cm]{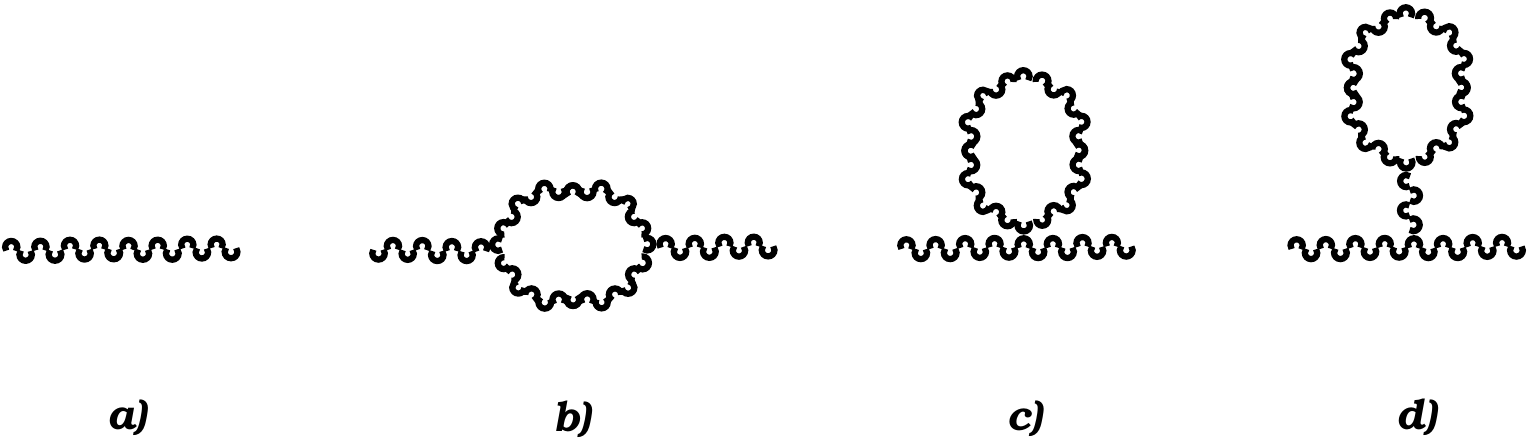} 
\caption{One-loop diagrams for the propagator $\LA{\vp}(z) \vp(0)\RA$.}
\label{fi:gen_pro}
\end{figure}

\begin{figure}
\includegraphics[width=13cm]{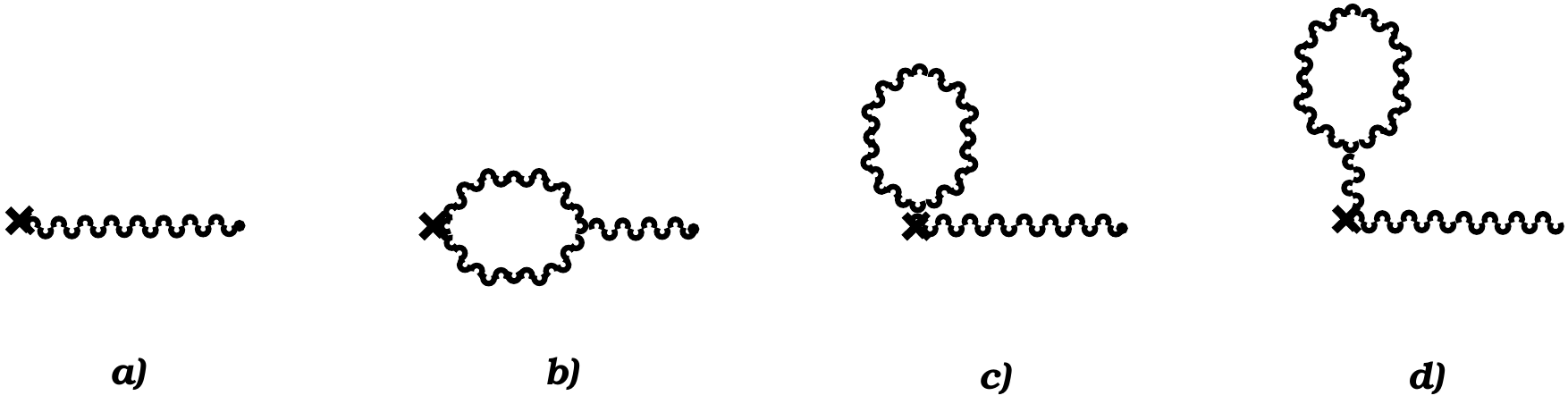} 
\caption{One-loop diagrams for $q/b^2$ in $T_{zz}^{({\rm eff})}$.}
\label{fi:gen_T1}
\end{figure}

In Ref.~\cite{Mak22} I computed $b^2$ and $q/b^2$ at the one-loop order 
(the first correction in $b_0^2$
to the classical values) by a straightforward evaluation of the associated  one-loop 
QFT diagrams shown in
Fig.~\ref{fi:gen_pro} and Fig.~\ref{fi:gen_T1}, respectively.
Both $b^2$ and $q/b^2$ look ugly for the given regularization and involve linear and logarithmic divergences, but the product
\be
\frac {q^2}{b^2} =\frac {q^2}{b^4} \times b^2=
\frac{1}{b_0^2} -\frac 16 -G +{\cal O}(b_0^2)
\label{100}
\ee
is simple and remarkably depends on $G$ in contrast to the usual DDK result for the Liouville action. 
The additional factor of $b^2$ in \eq{100}
is due to the  renormalization of $\vp$ by $b$.

\begin{figure}
\includegraphics[width=12cm]{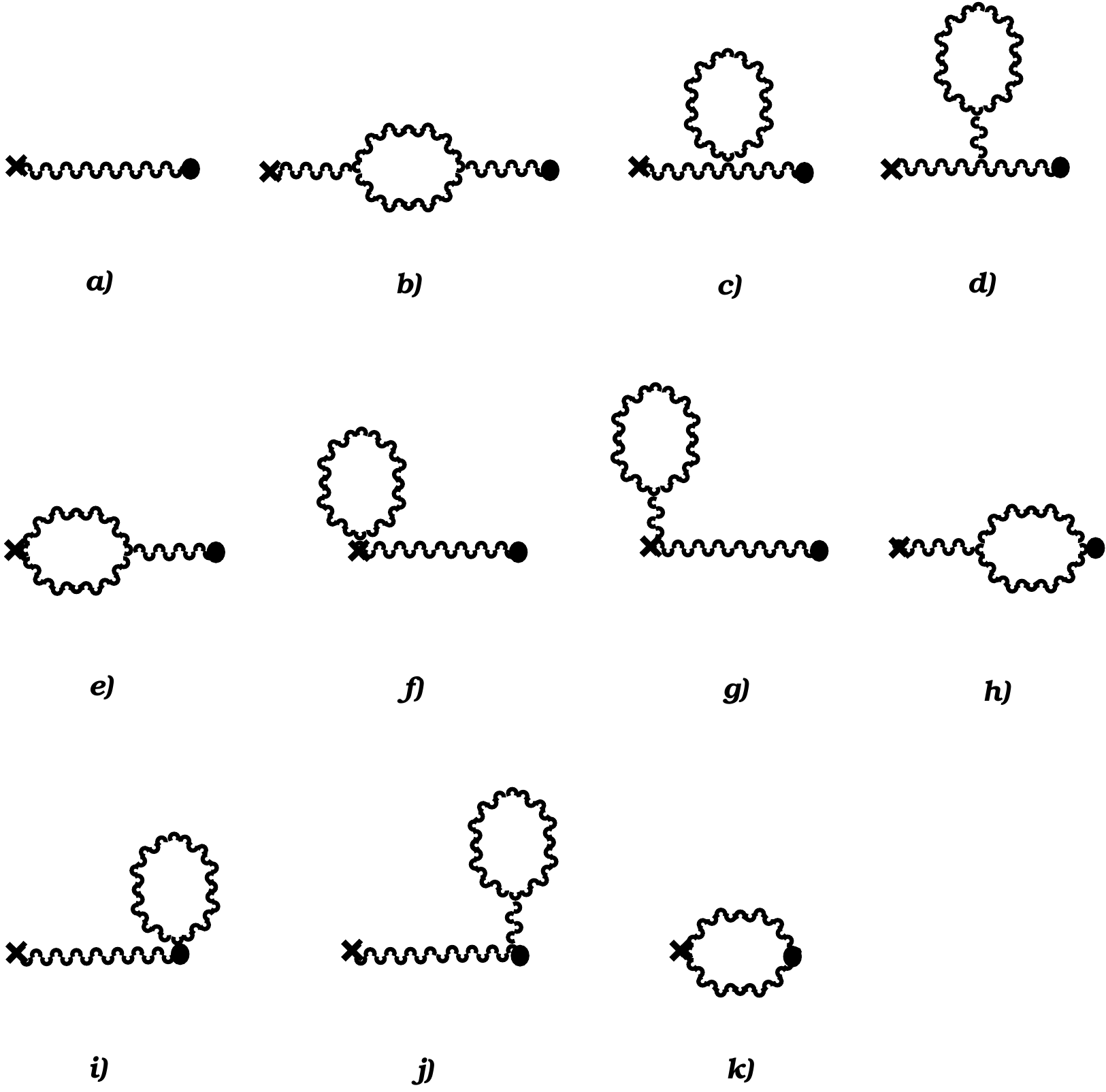} 
\caption{One-loop diagrams for the averaged operator product 
$\LA T_{zz}(z) T_{zz}(0)\RA$. The points $z$ and 0 are depicted 
by the cross and the dot, respectively.}
\label{fi:one-loop}
\end{figure}

In connection with the $d=1$ barrier 
%whose existence was found for the Polaykov string by Knizhnik-Polyakov-Zamolodchikov (KPZ)~\cite{KPZ}, 
it would be most interesting to understand how the DDK equation~\rf{DDK1} is to be modified
to all orders in $b_0^2$ for the action \rf{inva}.
An attempt to find out what happens at one loop is undertaken in Ref.~\cite{Mak22},
applying the methods of conformal field theory (CFT).
The computed averaged operator product reads
\be
\LA T_{zz}(z) T_{zz}(0) \RA =\frac c{2z^4},
\label{defc}
\ee
where $c$ is  the central charge  in the path-integral formalism.
It is given at one loop by the diagrams in Fig.~\ref{fi:one-loop}.
The diagrams a) to j) contribute $6 q^2/b^2$ to the central charge, while the diagram k) has
the finite part 
exactly coinciding with that in \eq{100} but also an additional logarithmic divergence
\be
c= \frac{6q^2}{b^2}+1+6 G \left(1-2\int \d k^2 \frac{\eps}{1+\eps k^2} \right)+{\cal O}(b^2_0).
\label{cphi}
\ee
Both  $\propto \!G$ finite and divergent parts come from the nonlocal (last) term in \rf{Tzz}.

I shall describe in these Notes how the logarithmic divergence of
 the central charge is canceled in the computation of a physical observable -- the string susceptibility. %simply ignored.
The reason for the appearance of that divergence 
are subtleties in the realization of conformal symmetry generated by the energy-momentum
tensor \rf{Tzz} which is not primary. I shall also confirm the final part in \rf{cphi} by
another technique.

These Notes are organized as follows. After a brief review of the setup in Sect.~\ref{s:2}
I describe in Sect.~\ref{s:3} massive conformal fields which can serve as 
Pauli-Villars' regulators preserving conformal symmetry. 
Sect.~\ref{s:cla} is devoted to some unusual properties of the
energy-momentum tensor~\rf{Tzz}. In Sect.~\ref{s:4} I introduce a new form
of the generator of conformal transformations based on the quantum equation of motion,
which drastically simplifies computations with the action~\rf{inva}. 
By computing the commutator of two conformal transformations  I show they
form the Lie algebra with a certain central extension. 
In Sect.~\ref{s:6} I compute the central charge at one loop by using the formulas 
derived in Sect.~\ref{s:4} and reproduce \eq{cphi} including the logarithmic divergence.
Finally, I demonstrate in Sect.~\ref{s:} how this  logarithmic divergence cancels in the
calculation of a physical observable -- the string susceptibility.
Some Conclusions are drown in Sect.~\ref{s:conclu}.
Appendix~\ref{appA} contains the Mathematica program demonstrating the conservation 
and tracelessness of the energy-momentum tensor ~\rf{Tzz}.
Appendix~\ref{appB} is devoted to the derivation of formulas for the singular products
involved in the computations.

\section{The setup\label{s:2}}    %% and a review of the previous results}

There are two ways of describing a quantum string: the Nambu-Goto or Polyakov formulations
which are commonly believed to be
equivalent except the dimension of the target-space 
coordinate $X^\mu$ is shifted
$d\to d-1$ because the metric $g_{ab}$ is independent for 
the Polyakov string  described by the  action
\be
S=\frac1{4\pi \alpha'}\int \sqrt{g} g^{ab} \p_a X \cdot \p_b X
\label{SPol}
\ee
which is quadratic in $X^\mu$ that makes it easy to integrate it out
in the path integral. Alternatively, the Nambu-Goto action is the area
of the string worldsheet which is highly nonlinear in $X^\mu$ but
can be made quadratic, introducing the Lagrange multiplier 
$\lambda^{ab}$ and an independent metric tensor $g_{ab}$,
\be
S_{\rm NG}=\frac1{2\pi \alpha'}\int \sqrt{\det {(\p_a X\cdot \p_b X)}}=
\frac1{2\pi \alpha'}\int \Big [\sqrt{g} +\frac 12\lambda^{ab} (\p_a X\cdot \p_b X-g_{ab}) \Big].
\label{SNG}
\ee

The equivalence of the two string formulations is evident at the classical level.
%It was also demonstrated~\cite{FTs82} at one loop about the classical string ground state.
The general argument in favor of the equivalence in the quantum case
is based~\cite{Pol87} on the fact that
$\lambda^{ab}$ propagates only to the distances of the order of the ultraviolet cutoff and
therefore should not affect the theory at large distances.
The statement relies on an effective action, governing fluctuations of $g_{ab}$ and 
$\lambda^{ab}$, which emerges after the path integration over
the target-space coordinates of the string (including possible Pauli-Villars' regulators) 
and the ghosts associated with fixing the conformal gauge \rf{confog}.
I shall not describe here this issue which is carefully reviewed in Ref.~\cite{Mak18}. 

Because the Lagrange multiplier $\lambda^{ab}$ does not ropagate to macroscopic distances 
it is tempting to
 additionally path-integrate over $\lambda^{ab}$ about the ground-state value 
 \mbox{$\bar {\lambda}^{ab}=\bar \lambda  \sqrt{g}  g^{ab}$} to obtain an effective action 
describing fluctuations of $g_{ab}$ in the infrared regime. 
Notice that $\blambda^{ab}$ does   not depend on $\vp$
 in the conformal gauge \rf{confog}. As was shown
in Ref.~\cite{Mak21}, we then arrive at the action \rf{inva} with $b_0^2=6/(26-d)$ and
a certain value of $G$ plus additional
terms of the order $\eps^2$ or higher. I did not mention herein the ground-state value of
the metric tensor $\bar g_{ab}=\bar \rho \hat g_{ab}$ all of which
are physically equivalent because of the so-called background
independence that stays in the given case that $\brho$ is not observable thanks to
the Weyl invariance.

A few comments concerning the derivation of the action \rf{inva} are in order.
If we set $\lambda^{z z}=\lambda^{\bar z \bar z}=0$ and path-integrate over
$\lambda^{z\bar z}$, then only the $R^2$ term in \rf{inva} would appear like
for the Polyakov string where it comes from the {Seeley} expansion of the heat kernel. 
The first term in \rf{inva} is associated with the
conformal anomaly while the $R^2$ term comes from the next order of the Seeley expansion 
in the worldsheet UV cutoff $\eps$.
It is familiar from the studies~\cite{KN93,Ich95,KSW} of the $R^2$ two-dimensional gravity,
where the first term in the action \rf{inva} was missing.
The second  higher-derivative term with $G\neq0$ emerges alternatively after 
integrating out $\lambda^{zz} $ and $\lambda^{\bar z \bar z}$.
The actual value of $G$  was not computed in~\cite{Mak21} but it was shown
to be nonvanishing for the Nambu-Goto string.  Only these two terms are independent
to order $\eps$. The others can be reduced to them modulo {boundary terms}
integrating by parts.

Of course the two higher-derivative terms  in the action~\rf{inva} are negligible classically for
smooth metrics as powers of $\eps R\ll1$, reproducing the Liouville action. 
However, the term with quartic derivative provides both a UV cutoff and also 
an {interaction} whose {coupling} constant is $\eps$.  We thus encounter 
 uncertainties like $\eps \times \eps^{-1}$ so the higher-derivative terms  {quantumly}  revive~\cite{Mak21} after doing these uncertainties. 
In other words  the  typical metrics which are essential in the
path integral over $\vp$ are not smooth and have $R\sim \eps^{-1}$. 
The first term in the action~\rf{inva} is thus
only an effective action governing smooth classical configuration, while 
the higher-derivative terms result in nontrivial interactions that cause
an eventful private life of $\vp$ which occurs at the very small distances of order $\sqrt{\eps}$
but nevertheless can be observable like the finite $G$ term in \eq{cphi}
(see \eq{gstrfin} below).
%All this is specific to the theory with diffeomorphism invariance.

There are also other higher-derivative terms of the order $\eps^2$ and higher 
additionally to \rf{inva} which are also not suppressed quantumly by $\eps\to0$.
One may expect  they do not change the results  owing to  
to the universality which often takes place near the critical point.
The arguments are presented in
Ref.~\cite{Mak21} where the universality was explicitly demonstrated for a six-order
high-derivative term $\sim \eps^2$. 
It looks like an appearance of anomalies in QFT.
This issue of yet  higher-derivative terms 
refers however to the strings as a whole
while the action~\rf{inva} by itself may be considered as a toy model of how to tell
the Nambu-Goto and Polyakov strings apart.

\section{Pauli-Villars regulators as massive conformal fields\label{s:3}}

To regularize divergences in the quantum case I
 implement the Pauli-Villars regularization,  adding to \rf{inva} 
 the following action for the regulator field $Y$:
\be
{\cal S}[Y] %^{({\rm reg})} 
=\frac 1{16\pi b_0^2}\int \sqrt{g}\left[ g^{ab}\partial_a  Y \partial_b Y +M^2 Y^2+
\eps (\Delta Y)^2 +G\eps g^{ab}\partial_a  Y \partial_b Y R  \right] .
\label{S0reg}
\ee
The field $Y$ has a very large mass $M$ and obeys wrong statistics to produce 
the minus sign for every loop, regularizing devergences coming from the loops of $\vp$.

To be precise, the introduction of one regulator is not enough to regularize all the divergences.
Some logarithmic divergences still remain. As was pointed out in \cite{AM17c}, the correct
procedure is to introduce two regulators of mass squared $M^2$ with wrong statistics, which can be viewed
as anticommuting Grassmann variables, and one regulator 
of mass squared $2M^2$ with normal statistics. Then all diagrams including quadratically divergent
tadpoles will be regularized. However, for the purposes of computing final parts just one regulator $Y$ would be enough. The contribution of the two others is canceled being mass independent.

The regulators also contribute to the energy-momentum tensor. The total one 
\be
T_{ab}=T_{ab}^{(\vp)}+T_{ab}^{(Y)}
\label{tttt}
\ee
is conserved and traceless
 thanks to the classical equations of motion for $\vp$ and $Y$
\bea
&&\hbox{l.h.s.\ side of \eq{cemG}}+\frac{M^2}2 Y^2-\frac\eps2 (\Delta Y)^2
+G\frac\eps 2 \p^a Y\p_a Y \Delta \vp -G\frac\eps 2\Delta (\p^a Y\p_a Y)
=0,\non &&
 \label{emphi} \\
&&-\Delta Y+M^2 Y +\eps \Delta^2 Y+G\eps \p_a (\p^a Y \Delta \vp ) =0
\label{emY},
\eea
respectively.
Thus the Pauli-Villars regulators are classically conformal fields in spite of they are massive.

This situation seems to be different from the usual one in QFT, 
where an anomaly emerges
if the regularization breaks the classical symmetry. We may thus expect that conformal
symmetry of the classical action \rf{inva} will be maintained  at the quantum level
for the Pauli-Villars regularization. 
%%owing  %%This is of cause due to diffeomorphism invariance. 
This can be confirmed  by explicit 
computations at one loop and partially  (for $\eps=0$) at two loops.

For the contribution of the regulator to $T_{zz}$ we find
\be
-4 b_0^2 T_{zz}^{(Y)}   %{({\rm reg})}
=\p Y \p Y -2\eps \p Y \p \Delta Y 
-G \eps \p Y\p Y \Delta \vp 
 + 4G\eps\p\vp \p(\e^{-\vp} \p Y \bp Y)  -4 G \q\eps \p^2 ( \e^{-\vp} \p Y \bp Y).
\label{Tzzreg}
 \ee

When I say ``regulators'' I mean large-mass fields with wrong statistics to provide minus signs for every loop.
But the above Eqs.~\rf{S0reg} to  %%, \rf{emphi}, \rf{emY}, 
\rf{Tzzreg} also apply to the case of a usual massive field with normal statistics  interacting with
two-dimensional gravity. Thus our consideration below also applies to such a model of the massive 
conformal field.

%%%%%%%%%%%%

%\section{Subtleties with OPE for non-quadratic {\boldmath{$T_{zz}$}}\label{appB}} 
\section{Properties of nonprimary energy-momentum tensor\label{s:cla}}

The usual definition of the central charge $c$ is linked to the transformation law
\be
\delta_\xi T_{zz} =\frac c{12} \xi''' + 2 \xi' T_{zz} +\xi \p T_{zz} 
\label{delTz}
\ee
of  the energy-momentum tensor  $T_{zz}$ under an  infinitesimal 
 conformal transformation $\delta z= \xi(z)$. 
 It is prescribed for the conserved tensorial primary
%\footnote{More precisely, $T_{zz}$  is the descendant of the primary unit operator.\label{foot}}
 field, which is the descendant of the primary unit operator, as was pointed out
in the original paper by Belavin-Polyakov-Zamolodchikov (BPZ) \cite{BPZ}.
Equation~\rf{delTz} is easily derivable for $T_{zz}$ which is
quadratic in $\vp$. 
%% and, as is stated in the footnote on p.~343 of \cite{BPZ}, correlators of $T_{zz}$ always
%% coincide with those of a quadratic $T_{zz}$  for free fields.
Let us check  how it works for  $T_{zz}$ given by \eq{Tzz} which is not quadratic in $\vp$.

It is easy to calculate how the energy-momentum tensor 
\rf{Tzz} changes under the  infinitesimal 
 conformal transformation. Substituting
\be
\delta_\xi \vp=\xi'+\xi \p \vp
\label{ctra}
\ee
into \rf{Tzz}, we find
\bea
\delta_\xi T_{zz} ^{(\vp)}&=&\frac 1{2 b_0^2} \xi''' + 2 \xi' T_{zz} ^{(\vp)}+\xi \p T_{zz} ^{(\vp)}+
\frac 1{b_0^2}G\eps \e^{-\vp}\Big\{ \xi'''' \bp \vp +\xi''' \big(\p\bp \vp-3 \p \vp \bp\vp\big) \non &&+
\xi''\Big[2 \bp\vp (\p\vp)^2-\p\vp\p\bp\vp-\bp\vp\p^2 \vp \Big]
-\e^\vp\frac1{\bp} \big[ \xi'' \p(\e^{-\vp} \bp\vp \p\bp\vp) \big]\Big\}.~~~~~
\label{delTza}
\eea
%or
%\bea
%\delta_\xi T_{zz} &=&\frac 1{2 b_0^2} \xi''' + 2 \xi' T_{zz} +\xi \p T_{zz} +
%\frac 1{b_0^2}G\eps \e^{-\vp}\Big\{ \xi'''' \bp \vp +\xi''' \big(\p\bp \vp-3 \p \vp \bp\vp\big) \non &&+
%\xi''\Big[2 \bp\vp (\p\vp)^2-\p\vp\p\bp\vp-\bp\vp\p^2 \vp \Big]
%+\e^\vp \frac1{\bp} \big( \xi'''\e^{-\vp} \bp\vp \p\bp\vp \big)\Big\}~~~~~
%\label{delTza}
%\eea
%\bea
%\delta_\xi T_{zz} &=&\frac 1{2 b_0^2} \xi''' + 2 \xi' T_{zz} +\xi \p T_{zz} +
%\frac 1{b_0^2}G\eps \e^{-\vp}\Big\{ \xi'''' \bp \vp +\xi''' \big(\p\bp \vp-3 \p \vp \bp\vp\big) \non &&+
%\xi''\Big[2 \bp\vp (\p\vp)^2-\p\vp\p\bp\vp-\bp\vp\p^2 \vp
%+\e^\vp \frac1{\bp} \e^{-\vp} \big(\p\vp \bp\vp \p\bp\vp-(\p\bp\vp)^2- \bp\vp \p^2\bp\vp \big)\Big]
%\Big\}~~~~~
%\label{delTza}
%\eea
%or
%\bea
%\delta_\xi T_{zz} &=&\frac 1{2 b_0^2} \xi''' + 2 \xi' T_{zz} +\xi \p T_{zz} +
%\frac 1{b_0^2}G\eps \e^{-\vp}\Big\{ \xi'''' \bp \vp
%+\xi''' \big(\p\bp \vp-3 \p \vp \bp\vp\big) \non &&+
%\xi''\Big[2 \bp\vp (\p\vp)^2-2\p\vp\p\bp\vp-\bp\vp\p^2 \vp \Big]\Big\}
%+\frac 1{b_0^2}G\eps \frac1{\bp} \Big[\xi''\e^{-\vp} \big(%%\pm 
%\p\vp \p\bp^2 \vp- \bp\vp \p^2\bp\vp \big)\Big].~~~~~
%\label{delTza}
%\eea
It deviates from \eq{delTz}  by the presence of the additional terms which
arise because the action \rf{inva} involves
the structure 
\be
L= g^{ab}\p_a\vp \p_b\vp =4\e^{-\vp} \p \vp \bp \vp
\label{defL}
\ee
(familiar from the Lagrangian of a free field)
which is scalar but not primary and 
transforms %%under \rf{ctra}
as
\be
\delta_\xi L = \xi \p L +
4\xi'' \e^{-\vp} \bp\vp.
\ee
%\be
%\delta_\xi \left(\e^{-\vp} \p\vp \bp\vp\right) = \xi \p \left(\e^{-\vp} \p\vp \bp\vp\right) +\xi'' \e^{-\vp} \bp\vp.
%\ee

The additional terms %%in \rf{delTza} 
do not appear for $G=0$ because
the scalar curvature {$R=-4 \e^{-\vp}\p\bp \vp$} is a primary scalar:
\be
\delta_\xi R = \xi \p R.
\ee
Then the action is not quadratic in $\vp$ but the usual CFT technique perfectly works.

It is also easy to repeat the calculation for the total energy-momentum tensor
\rf{tttt} which is the
sum of \rf{Tzz} and that of the regulators \rf{Tzzreg}. %Denoting it  by $T_{zz}$, 
Accounting for the
transformation of $Y$ as a primary scalar
\be
\delta_\xi Y = \xi \p Y
\ee
and noting that
\be
\delta_\xi T_{zz} ^{(Y)}= 2 \xi' T_{zz} ^{(Y)}+\xi \p T_{zz} ^{(Y)},
\label{22}
\ee
we arrive again at \eq{delTza} with $T_{zz}^{(\vp)}$ substituted by the total $T_{zz}$.

The additional terms in \rf{delTza} vanish as $\eps\to0$, so they do not affect then the classical limit if
$\vp$ is smooth. However,
averaging \rf{delTza} over $\vp$ with the quadratic weight, we get the following $\xi'''$ term:
\be
\LA\delta_\xi T_{zz}^{(\vp)}(0)\RA = \xi''' (0)\left( \frac 1{2 b_0^2} - G \int \d k^2 \frac {\eps}{1+\eps k^2}
\right),
\label{xi3}
\ee
where we substituted
\be
\LA \eps L \RA = \LA \eps R \RA =2\int \d k^2 \frac {\eps}{1+\eps k^2}.
\label{25}
\ee
All other terms in \rf{delTza} do not contribute to the average.
The  logarithmically divergent second term in \rf{xi3} is just the same as in \eq{cphi} what may help to understand
its appearance there. While in the calculation of \rf{cphi} in Ref.~\cite{Mak22} I used $\eps\to0$,
it has been now derived for an arbitrary $\eps$.
It is worth noting that the logarithmic divergence would not appear in the operator formalism where 
the operators %%%in $T_{zz}$ 
are normal-ordered and the vacuum expectation value $\LA0 | \!: \!\!L \!\!:\!| 0\RA=0$.

%As  concerning the last (nonlocal) term in \eq{delTza}, one may naively think that it vanishes but there are
%subtleties because the associated integrals again diverge, so an uncertainty like $0 \times \infty$ emerges.
%I shall now describe the technique how to deal with it.

Several questions immediately arise as to the definition of the central charge, which is linked 
to the Virasoro algebra, for a nonprimary energy-momentum tensor. 
I shall answer part of them in the next two sections.
%% but hope to return to this issue elsewhere.

\section{Conformal symmetry at the quantum level\label{s:4}}

For a general action $S[\vp]$ the generator of the conformal transformation can be written as
\be
\hat \delta_\xi \equiv \int_{C_1}\frac{\d z}{2\pi \i} \xi (z) T_{zz} (z)=\frac1\pi \int _{D_1}\xi \bp T_{zz}\stackrel{{\rm w.s.}}
=\int_{D_1} \left(\q \xi' \frac {\delta}{\delta \vp}
 +\xi \p \vp\frac {\delta }{\delta \vp}\right)
 \label{hatdel}
 \ee
 with
 \be
 \frac{\delta \vp(0)}{\delta \vp(z)} =\delta^{(2 )}(z),
 \qquad \delta^{(2 )}(z)=\bp \frac1{\pi z}.
 \label{defdel}
\ee
We have used the (quantum) equation of motion
\be
\bp T_{zz} =-\pi \q \p \frac {\delta S}{\delta \vp} + \pi \p \vp\frac {\delta S}{\delta \vp},\qquad
\frac {\delta S}{\delta \vp} \stackrel{{\rm w.s.}}=\frac {\delta}{\delta \vp}
\label{26}
\ee
%%with $\qq=1$ 
and integrated by parts.
The domain $D_1$ includes the singularities of $\xi(z)$ leaving outside possible singularities of the function $X(\om_i)$ 
on which \rf{defdel} acts and $C_1$ bounds $D_1$. In the classical limit $b_0^2\to 0$ we reproduce \rf{ctra}
by acting with the right-hand side of \eq{hatdel} on $\vp$.

An advantage of the proposed form of $\hat \delta_\xi$ on the right-hand side of \eq{hatdel} over the standard
one on the left is that it takes into account a tremendous cancellation of the diagrams in the quantum case we
now proceed, while there are subtleties associated with singular products. The results of computing the variational 
derivative may seem to be the same as in a free theory (with the quadratic action) but their averages can differ.
I would classify the calculation below as a pragmatic mixture of
the  CFT and QFT methods.

In the quantum case we have an additional effect of the regulator
\be
\LA\hat \delta_\xi X(\om_i) \RA =\LA \int_{D_1} \d^2 z \left(\q \xi' (z)\frac {\delta}{\delta \vp(z)}
 +\xi(z)\p \vp(z)\frac {\delta }{\delta \vp(z)}+\xi(z)\p Y(z)\frac {\delta }{\delta Y(z)}\right)X(\om_i) \RA.
 \label{XX}
\ee
%%%Each variational derivative  adds $b_0^2$ %%on the right-hand side of 
%%%according to the quantum 
%%%equation of motion \rf{SDe}, which makes \rf{XX} convenient for the loop expansion.
Averaging over the regulators, we arrive at the effective action, governing fluctuations of $\vp$,
and the effective energy-momentum tensor, 
which in the infrared limit becomes quadratic in $\vp$ as shown in \eq{Teff}. The arguments are similar to DDK.
Equation~\rf{XX} is then substituted by
\be
\LA\hat \delta_\xi X(\om_i) \RA =\LA \int_{D_1} \d^2 z \left(\qq \xi' (z)\frac {\delta}{\delta \vp(z)}
 +\xi(z)\p \vp(z)\frac {\delta }{\delta \vp(z)}\right)X(\om_i) \RA.
 \label{XXq}
\ee
As is demonstrated in Ref.~\cite{Mak22} by explicit calculations at one loop, these two ways
of computing the central charge are complementary: we either add at $q=1$ the contributions from $\vp$  and the regulators or consider at $q\neq 1$ only the contribution from $\vp$ \`a la DDK.

Given \eq{XXq} it is instructive to reproduce
\be
\hat \delta_\xi \e^{\vp(\om)} \stackrel{{\rm w.s.}}=(q-b^2)\xi'(\om)\e^{\vp(\om)} +\xi(\om)\p\vp(\om) \e^{\vp(\om)}
\label{29}
\ee
 for the quadratic action. We obtain
 \bea
 \lefteqn{
\LA \hat \delta_\xi \e^{\vp(\om)} X\RA=\int_{D_1}\!\d^2 z 
\LA\big[ q\xi'(z)+\xi(z) \p \vp(z) \big] \frac\delta{\delta \vp(z)}\e^{\vp(\om)} X\RA } \non  &=&
q \xi'(\om) \LA  \e^{\vp(\om)} X\RA +\int_{D_1}\!\d^2 z \, \xi(z) 
\LA  \p\vp(z)\e^{\vp(\om)} X\RA \delta^{(2)}(z-\om)\nonumber \\ &=&
q \xi'(\om) \LA  \e^{\vp(\om)} X\RA +\int_{D_1}\!\d^2 z \, \xi(z) 
\big\langle  \p\vp(z) \vp(\om)\big\rangle \delta^{(2)}(z-\om) \LA \e^{\vp(\om)} X\RA\non &&
%% \hspace*{1cm}
 + \xi(\om) \LA  \p\vp(\om)\e^{\vp(\om)} X\RA. % \non
\label{32}
 \eea
 The most interesting is the second term on the right-hand side, where the singular product equals
\be 
\int_{D_1}\!\d^2 z \, \xi(z) 
\big\langle  \p\vp(z) \vp(\om)\big\rangle \delta^{(2)}(z-\om) =-b^2 \xi'(\om)
\label{most}
\ee
as shown in \eq{C7} of Appendix~\ref{appB},
reproducing \rf{29}.

If we repeat this computation for the case of the higher-derivative action~\rf{inva}, we still 
infer  the second line in \eq{32} from the first one but now the average in the second line does not
factorize in general because of the interaction, unless the diagrams with interaction are mutually
canceled. Nevertheless, the factorization holds at the one-loop order where the passage from the
second to the third line in \eq{32} works and the averages are to be calculated in the non-interacting
 higher-derivative theory. 
We thus  obtain to order $b_0^2$ the same conformal weight of $\e^\vp$
as in \rf{29} because \eq{most} still holds in this case as shown in Appendix~\ref{appB}.

Using \eq{XXq} we have for the commutator of two conformal transformations 
\bea
\lefteqn{\LA(\hat \delta_\eta \hat \delta_\xi -\hat \delta_\xi \hat \delta_\eta) X \RA =
\LA \hat \delta_\zeta  X \RA }\non &&\hspace*{1cm}+ \int_{D_1}\!\d^2 z \int_{D_z} \!\d^2 \om
\LA\big[ q\xi'(z)+\xi(z) \p \vp(z) \big] \big[ q\eta'(\om)+\eta(\om) \p \vp(\om) \big]
\frac{\delta^{2} S}{\delta\vp(z) \delta \vp(\om)} X\RA \non &&
\label{Viraso0}
\eea
with $\zeta= \xi\eta'-\xi'\eta$ as it should. Here the domain $D_1$  includes the singularities of 
$\xi(z)$ and $\eta(z)$, 
leaving outside possible singularities of $X$,
and $D_z$  comprises $z$. This represents (see e.g.~\cite{Pol87}) the commutator in the operator formalism.
The first term on the right-hand side is linked to the classical \eq{hatdel}
while the second term can be written in the form
\be
\LA(\hat \delta_\eta \hat \delta_\xi -\hat \delta_\xi \hat \delta_\eta) X \RA=
\LA \hat \delta_\zeta  X \RA+ \frac{1}{24} \oint _{C _1}\frac{\d z}{2\pi \i} \big[ \xi'''(z)\eta(z)-\xi(z)\eta'''(z) \big]
\Big\langle cX \Big\rangle
\label{Viraso}
\ee
which is usually linked to the central charge of the Virasoro algebra. 
As we shall momentarily see, $c$ in \eq{Viraso}
is the usual c-number for the quadratic action but it
will be $\vp$-dependent for the higher-derivative action \rf{inva} with $G\neq 0$, 
as we observed already for the term with $\xi'''$ in \eq{delTza}.

\section{The central charge at one loop\label{s:6}}

Let us begin by showing how to reproduce from \eq{Viraso} the usual results for  the quadratic action.
Noting that
\be
\frac{\delta^{2} S}{\delta\vp(z) \delta \vp(\om)}=\frac{1}{8\pi b^2} (-4\p\bp) \delta^{(2)}(z-\om)
\ee 
for the quadratic action, we find
\bea
 \int_{D_1}\! \!\d^2 z \int_{D_z} \!\!\d^2 \om \,
q^2\xi'(z)\eta'(\om)
\frac{\delta^{2} S}{\delta\vp(z) \delta \vp(\om)}
=\frac{q^2}{4b^2} \oint _{C _1}\frac{\d z}{2\pi \i} \big[ \xi'''(z)\eta(z)-\xi(z)\eta'''(z) \big]
\label{c21}
\eea
and
\bea
 \lefteqn{\int_{D_1}\! \!\d^2 z \int_{D_z} \!\!\d^2 \om \,\xi(z)\eta(\om)
\LA \p \vp(z)  \p \vp(\om) \RA
\frac{\delta^{2} S}{\delta\vp(z) \delta \vp(\om)} }\non
&&=-\frac 1{2b^2} \oint _{C _1}\frac{\d z}{2\pi \i} \xi(z) \int _{D_z}\d^2\om
 \big[ \eta'(z)\LA \p^2\vp(\om) \vp(z) \RA
+\eta(z)\LA \p^3\vp(\om) \vp(z) \RA\big] 
\non
&&=\frac{1}{24} \oint _{C _1}\frac{\d z}{2\pi \i} \big[ \xi'''(\om)\eta(z)-\xi(\om)\eta'''(z) \big] ,  %%\,\frac 16.
\label{c22}
\eea
where we have used the formula
\bea
%\int \d^2 z\, \xi(z) \LA \partial^n \vp(z) \vp(0) \RA 
\int_{D_z} \d^2 \om\, f(\om)  \LA \partial^n \vp(\om) \vp(z)  \RA
\delta^{(2)}(\om-z) =(-1)^{n}  b^2 H_n f^{(n)}(z) ,
\hspace*{.1cm}\non
H_1=1,\qquad H_2=\frac13,\qquad H_3=\frac 16,\qquad H_n=\frac{2}{n(n+1)}~~~
%%%C_4=\frac1{10},~~C_5=\frac1{15},~~C_6=\frac1{21},~~C_7=\frac1{28}
\label{C7vp}
\eea
for the singular products in CFT with the quadratic action.  It generalizes \eq{most} and is derived in Appendix~\ref{appB}.
The sum of \rf{c21} and \rf{c22} gives  $c=6 q^2 /b^2+1$, reproducing    DDK.

It is more lengthy to deal with the higher-derivative terms in the action
\rf{inva} which we expand in the powers of $\vp$ as
\be
{\cal S}={\cal S}^{(2)}+{\cal S}^{(3)}+{\cal S}^{(4)}+{\cal O}(\vp^5)
\ee
with
\begin{subequations}
\bea
{\cal S}^{(2)}&=&\frac 1{4\pi b^2} \int \Big[ \p\vp \bp \vp +4 \eps (\p\bp \vp)^2 \Big],
 \label{S2}\\
{\cal S}^{(3)}&=&-\frac 1{\pi b^2} \int \Big[ \eps \vp (\p\bp \vp)^2 
+G\eps \p \vp \bp \vp \p\bp \vp\Big], \label{S3}\\
{\cal S}^{(4)}&=&\frac 1{\pi b^2} \int \Big[ \frac 12 \eps \vp^2 (\p\bp \vp)^2 
+G\eps \vp\p \vp \bp \vp \p\bp \vp\Big].
\label{S4}
\eea
\label{S123}
\end{subequations}
The next orders will not contribute at one loop. 

The average in the second term on the right-hand side of \eq{Viraso0} factorizes 
at one loop and we obtain for the nonvanishing terms with $G$
\bea
 &&\int_{D_1}\!\d^2 z \int_{D_z} \!\d^2 \om \,
\xi'(z)\eta'(\om) \LA
\frac{\delta^{2} S^{(4)}}{\delta\vp(z) \delta \vp(\om)} \RA
=\frac{G\eps}{b^2}\int \d^2 z \d^2\om \d^2 t \, \xi'(z)\eta'(\om) \LA  \p\vp(t)\bp\vp(t)\RA
\non && \hspace*{4cm}\times\big[  \delta^{(2)}(t-z) \p\bp \delta^{(2)}(t-\om) 
+ \p\bp\delta^{(2)}(t-z)  \delta^{(2)}(t-\om) \big]
\non %\hspace*{1cm}
&&\hspace*{1cm}=\frac{1}{4} \oint _{C _1}\frac{\d z}{2\pi \i} \big[ \xi'''(z)\eta(z)-\xi(z)\eta'''(z) \big]
\left(  -2 G \int \frac{\eps \d k^2}{1+\eps k^2}  \right)
\label{ic21}
\eea
and
\bea
 \lefteqn{\int_{D_1}\!\d^2 z \int_{D_z} \!\d^2 \om \,
\LA\big[\xi'(z)\eta(\om) \p\vp(\om)+ \xi (z)\eta'(\om) \p\vp(z)\big]
\frac{\delta^{2} S^{(3)}}{\delta\vp(z) \delta \vp(\om)} \RA}\non
&&=- \frac{G\eps}{b^2}
\int \d^2 z \d^2\om \d^2 t \LA \big[ \xi'(z)\eta(\om)\p\vp(\om)+ \xi(z)\eta'(\om)\p\vp(z)\big] \p\bp\vp(t)\RA \non
&& ~~~~~~~\times
\big[  \p\delta^{(2)}(t-z) \bp \delta^{(2)}(t-\om) + \bp\delta^{(2)}(t-z) \p \delta^{(2)}(t-\om) \big]
\nonumber \\&&=\frac{G\eps}{b^2}  \oint _{C _1}\frac{\d z}{2\pi \i} 
\int_{D_z} \d^2 \om \big[ \xi''(z)\eta(\om)+\xi(z)\eta''(\om) \big] 
\LA \p^2\bp \vp(\om)\vp(z) \RA \delta^{(2)} (\om-z)
\non %%\hspace*{1cm}
&&=\frac{1}{4} \oint _{C _1}\frac{\d z}{2\pi \i} \big[ \xi'''(z)\eta(z)-\xi(z)\eta'''(z) \big]
\, G  .
\label{ic22}
\eea
Here we have used \eq{25} and substituted
\be
 \int_{D_z} \d^2 \om \,  f(z)\LA -4\eps\p^2\bp \vp(\om)\vp(z) \RA \delta^{(2)} (\om-z) =-b^2 f'(z)
 \label{44}
\ee
as calculated in Appendix~\ref{appB} for the propagator in a higher-derivative theory
which is just what we have in \eq{44} in the one-loop approximation.

The sum of \rf{ic21} and \rf{ic22} remarkably reproduces the 
$\propto\!\!G$ addition to the central charge in \eq{cphi}.
I also mention that the same final part can be obtained additionally to \rf{xi3} in a slightly different way 
by a direct  computation of $\LA \hat \delta T_{zz}^{(\vp)}\RA $ at one loop.

%%In the next section I shall do that in a slightly different way with more easy computations.

%\section{Renormalization of the central charge\label{s:}}
\section{Cancellation of the logarithmic divergence\label{s:}}

For the usual Polyakov string the central charge of $\vp$ is observable, e.g.\ via the string susceptibility index
$\gamma_{\rm str}$. I show in this section that it remains finite at one loop for the higher-derivative
action~\rf{inva} in spite of the logarithmic divergence of the central charge which cancels in
$\gamma_{\rm str}$.
%% noan additional renormalization is required

The gauge \rf{confog} implies the relation
\be
\sqrt g R \Rightarrow  \sqrt{\hat g} \left( q\hat R - \hat \Delta \vp \right)
\label{Rshift}
\ee
between the scalar curvatures $R$ and $\hat R$ for 
the metrics $g_{ab}$ and $\hat g_{ab}$, respectively, and analogously
\be
\Delta\equiv\frac 1 {\sqrt{ g}}  \partial_a \sqrt{ g}  g^{ab} \partial _b
=\e^{-\vp} \frac 1 {\sqrt{\hat g}} \partial_a \sqrt{\hat g} \hat g^{ab} \partial _b \equiv
\e^{-\vp} \hat \Delta
\label{defDel}
\ee
between  the two-dimensional Laplacians. The factor $q$ is as before to appear in \eq{Teff}.

The string susceptibility index can be derived~\cite{DDK} from the response of the system to
 a uniform dilatation of space, which means adding a constant to $\vp$. 
 The quadratic part of the action then results in  the topological Gauss-Bonnet term producing
the Euler characteristic $2-2h$ in %% the string susceptibility index 
\be
%%\gamma_{\rm str}=(1-h)\frac q{ b^2} +\gamma_1
\gamma_{\rm str}=(1-h)\frac {q}{b^2 }+\gamma_1=(1-h)\frac {q^2}{2 b^2}\left(   
1+\sqrt{1-\frac{4b^2}{q^2}}\right) +\gamma_1
\label{gstr}
\ee
where $\gamma_1=2$ for a closed string in $R_d$. This
reproduces the usual $\gamma_{\rm str}$~\rf{ggg} of KPZ-DDK.
 
For the higher-derivative action~\rf{inva}  we have an additional contribution 
from the term with $G$ which reads
\be
 {\cal S}_G=
-\frac{G \eps}{16\pi b^2} \int \e^{-\vp} \sqrt{\hat g} \hat g^{ab} \partial_a 
\Big( \vp-\frac q{\hat \Delta}\hat R \Big)\partial_b 
\Big( \vp-\frac q{\hat \Delta}\hat R \Big) \big(\hat \Delta \vp - q\hat R\big).
\ee
The relation \rf{Rshift} is used in the derivation.
 While this term is negligible for smooth configurations as $\eps\to0$, it revives after averaging over $\vp$ as is already explained.
 
For a constant infinitesimal $\delta \vp$ the variation of the action is expressed via the left-hand
 side of the classical equation of motion~\rf{cemG} where we substitute
$
 \vp\Rightarrow \vp+q \hat \vp
 $
 to account for the conformal background metric. The first term on the left-hand
 side of \eq{cemG} immediately gives the Gauss-Bonnet term. It is also easy to show the cancellation
 at one loop between the third and fourth terms which are present at $G=0$. 
 Therefore, they do not contribute to  $\gamma_{\rm str}$.

Concerning the terms with $G\neq0$, only the last one in \eq{cemG}
gives a nonvanishing  contribution at one loop after averaging
over $\vp$. We find 
 \be
 \LA\delta {\cal S}_G \RA=-
\frac{G q\eps}{8\pi b^2} \delta \vp \int  \sqrt{\hat g} \hat g^{ab} 
\LA \e^{-\vp}\partial_a \vp \partial_b \vp\RA \hat R 
\label{48} 
\ee
which gives an addition to the right-hand side of \eq{gstr}.
Expanding to the one-loop order where our results coincide
with DDK except for  the additional terms in \rf{cphi} at $G\neq0$, we find
%\be
%\gamma_{\rm str}=
%(1-h)\left[\frac {q^2}{2 b^2}\Big(   
%1+\sqrt{1-\frac{4b^2}{q^2}}\Big) -2Gq\int \d k^2\frac{\eps }{1+\eps k^2}\right] +\gamma_1.
%\label{gstrL}
%\ee
\be
\gamma_{\rm str}=
(1-h)\left[\frac {q^2}{b^2}-1 -2G\int \d k^2\frac{\eps }{1+\eps k^2} +{\cal O}(b_0^2)\right] +
\gamma_1.
\label{gstrL}
\ee
Extracting $q^2/b^2$ from \eq{cphi},
we see the cancellation between
the logarithmic divergence in \rf{gstrL} and the logarithmic divergence in the central charge. The contribution
of the final part in \rf{cphi} remains and we obtain from \rf{gstrL} 
\be
\gamma_{\rm str}=
(1-h)\left( \frac {1}{b_0^2}  -\frac76-G +{\cal O}(b_0^2)\right) +\gamma_1
\label{gstrfin}
\ee
showing for $G\neq0$ a deviation from the one-loop result~\cite{Zam82,CKT86}
for the Polyakov string in $d$ target-space dimensions for which $b_0^2=6/(26-d)$ and $G=0$,
thus confirming the results of Ref.~\cite{Mak22}.
In the operator formalism the logarithmic divergence does not appear both in the central charge, as is already mentioned in Sect.~\ref{s:cla},
and in \eq{gstrL} because of the normal ordering of the operators in $S_G$, resulting in the vanishing of
the vacuum expectation value $\LA0 | \!:\! \!L \!\!:\!| 0\RA$ in \eq{48}.

\section{Conclusion\label{s:conclu}}

As I already mentioned in the Introduction, my main motivation for these Notes 
was to confirm the results of Ref.~\cite{Mak22} and to understand the
origin of the logarithmic divergence in the one-loop central charge of $\vp$ for the conformal
theory with the higher-derivative action~\rf{inva} at $G\neq0$. This happens because its energy-momentum tensor~\rf{Tzz}
is not  the descendant of the primary unit operator
%%a primary (see footnote~{\ref{foot}}) tensor 
owing to the presence of the structure~\rf{defL} which is not a primary scalar.
Nevertheless, this logarithmic divergence is what a doctor ordered to have finite $\gamma_{\rm str}$ 
which would diverge otherwise in the path-integral formalism.

I have developed a kind of the general technique to deal with the conformal transformation generated by 
a nonprimary energy-momentum tensor. It is based on using the representation \rf{hatdel} 
of the generator, which takes into account the quantum equation of motion and involves singular 
products in the quantum case. Using this technique, I have calculated the commutator of two 
conformal transformations with  the central extension shown in \eq{Viraso0}. For the quadratic 
action it reproduces the usual results, while for the higher-derivative action~\rf{inva}
the central extension being $\vp$-dependent  %% and its vacuum average produces the logarithmic divergence.
still results at one loop in the Virasoro algebra with the central charge given by the sum of
\rf{ic21} and \rf{ic22}.

The most important task will be of course to go beyond the one-loop approximation for the
 higher-derivative action~\rf{inva} to understand how  $\gamma_{\rm str}$ may depend on
 $d$ for the Nambu-Goto string represented by $G\neq 0$ in our toy model. 
 A very interesting problem will be to find  out 
 what Mathematical structures may be encoded
 in \rf{Viraso} with a $\vp$-dependent $c$.
This is beyond the subject of these Notes dealing with the one-loop order.
%whose goal was to understand the origin of the logarithmic divergence
%in \rf{cphi} and to confirm the results of Ref.~\cite{Mak22}.

\subsection*{Acknowledgement}

I am grateful to the Referee for bringing to my attention two references not mentioned in
these Notes.

This work was supported by the Russian Science Foundation (Grant No.20-12-00195).

%\pagebreak
\appendix

\section{Mathematica program showing conservation and tracelessness of $\boldsymbol{T_{ab}}$\label{appA}}

This Mathematica program 
can be copied and pasted in Mathematica to demonstrate
the conservation and tracelessness of the energy-momentum tensor~\rf{Tzz}.

\begin{verbatim}

(* checking the conservation and traceless of T_ab for any G *)
(* the Liouville field *)
v = vp[z, bz]
G =.
(* Tzz for minimal interaction *)
Tmzz = D[v, 
     z]^2 + (1 - G) (-2 eps D[v, z] D[4 Exp[-v] D[v, z, bz], z]) + 
   4 eps G (D[v, z] D[Exp[-v], z, z, bz] + 
      D[Exp[-v], z] D[v, z, z, bz]);
(* minimal T_ab is not traceless *)
Tmzbz = -4 G eps D[v, z, bz] D[Exp[-v], z, bz] + 
   4 (1 - G) eps Exp[-v] D[v, z, bz]^2;
Expand[%]
(* classical equation of motion -> cem=0 *)
cem = -D[v, z, bz] + 
   eps  (1 - G) (D[4 Exp[-v] D[v, z, bz], z, bz] - 
      1/2 Exp[-v] 4 D[v, z, bz]^2) + 
   2 eps G (Exp[-v] D[v, z, z, bz, bz] - D[Exp[-v], z, z, bz, bz]);
(* conservation of minimal Tzz *)
Expand[D[Tmzz, bz] + D[Tmzbz, z] + 2 D[v, z] cem]
(* contribution from nonminimal interaction except nonlocal term *)
delTzz = -2 (1 - G) D[v - 4 eps Exp[-v] D[v, z, bz], z, z] - 
   2 G D[v - 4 eps Exp[-v] D[v, z, bz] + 
      2 eps Exp[-v] D[v, z] D[v, bz], z, z] + 
   4 eps  G D[Exp[-v] D[v, z, bz] D[v, z], z];
(* additional contribution from nonlocal term *)
addi = eps G D[D[v, bz] 4 Exp[-v] D[v, z, bz], z, z];
(* final calculation of d_bz T_zz *)
Expand[D[Tmzz + delTzz, bz] + addi + 2 D[v, z] cem - 2 D[cem, z]]
(* =0 -> it works! *)

\end{verbatim}

%\pagebreak
\section{Web of formulas for the singular products\label{appB}}

Let me begin by mentioning a very important role played in CFT by the formula
\be
\delta^{(2)}(z) =\bp \frac {1}{\pi z}.
\label{very}
\ee
Its big brothers can be derived using the obvious identity
\be
\frac 1{z^n} \bp \frac 1z=(-1)^n \frac1{(n+1)!} \p^n \bp \frac 1z.
\label{B2}
\ee
Proceeding this way, we arrive at the formula
\bea
%\int \d^2 z\, \xi(z) \LA \partial^n \vp(z) \vp(0) \RA 
8\pi \int \d^2 z\, \xi(z)  \partial^n G_0(z)  
\delta^{(2)}(z) =(-1)^{n}  H_n \xi^{(n)}(0) ,
\hspace*{1cm}\non
H_1=1,\qquad H_2=\frac13,\qquad H_3=\frac 16,\qquad H_n=\frac{2}{n(n+1)}.
%%%C_4=\frac1{10},~~C_5=\frac1{15},~~C_6=\frac1{21},~~C_7=\frac1{28}
\label{C7}
\eea
for the singular products in a free CFT 
with the propagator
\be
G_0(z) =-\frac1{2\pi} \log\Big(\sqrt{z\bz} \mu\Big),
\label{G0}
\ee
where $\mu$ represents a IR cutoff. 

I derive in this Appendix \eq{C7} and other similar formulas explicitly taking into account a regularization
 of the singular products.
 
 %%\subsection{Quadratic action}

To derive  \rf{C7} and its generalizations, we
 introduce a  UV regularization $a$ for example by 
\be
G_a(z)=-\frac 1{4\pi} \log \big[(z\bz+a^2)\mu^2 \big]
\label{simplest}
\ee
and regularize the delta function by
\be
\delta ^{(2)}_a(z) =-4 \p\bp G_a(z)=\frac{a^2}{\pi(z\bz+a^2)^2}
\ee
as is prescribed by the formula of the type
\be
-\frac1{2b_0^2 \pi}\int \d^2 z\, \xi(z) \LA \big( \p\vp(z) \p\bp\vp(z) \big) \e^{\vp(0) }\RA 
=8\pi b_0^2 \int \d^2 z\, \xi(z) \p G_a(z) (-4\p\bp) G_a(z)  \LA\e^{\vp(0) }\RA 
\label{Cccc}
\ee
for the conformal transformation of a free field.
We then derive
\bea
8\pi 
\int \d^2 z\, \xi(z) \p^n G_a(z) \delta _{a}^{(2)}(z)=(-1)^{n}  H_n \xi^{(n)}(0) ,
\label{C7a}
\eea
reproducing \eq{C7} in the limit $a\to0$.

It is instructive to repeat the computation for the proper-time regularization, where the lower 
limit of the integrals over the
proper time is $\tau>0$. We then have
\be
G_\tau(z)=\frac1{4\pi} \left[  {\rm Ei}\Big(-\frac{z \bz}{4\tau} \Big)-\log(z\bz \mu^2)\right],
\qquad
\delta^{2}_\tau(z) =-4\p\bp G_\eps(z)=\frac{\e^{-z\bz/4\tau}}{4\pi \tau}
\label{Gtau}
\ee
with Ei being the exponential integral,   %%and
%\be
%\delta^{2}_\tau(z) =-4\p\bp G_\eps(z)=\frac{\e^{-z\bz/4\tau}}{4\pi \tau}
%\ee
which gives
\bea
8\pi 
\int \d^2 z\, \xi(z) \p^n G_\tau(z) \delta _{\tau}^{(2)}(z)=(-1)^{n}  \frac 1{2^{n-1}n} \xi^{(n)}(0) .
\label{C7tau1}
\eea
%and
%\bea
%8\pi \int \d^2 z\, \xi(z) (-4\tau\p^{n+1}\bp) G_\tau(z) \delta _{\tau}^{(2)}(z)=(-1)^{n}  \frac 1{2^{n}} \xi^{(n)}(0) .
%\label{C7tau2}
%\eea

Rather surprisingly the numbers are not universal except for $n=1$
and depend on the regularization applied. However, 
for the central charge in \eq{c22} we have %% from \eq{57}
\be
\frac 12\left( \frac 13-\frac 16 \right)=\frac {1}{12}
\ee
from \eq{C7a} and
\be
\frac 12\left( \frac 14-\frac 1{12} \right)=\frac {1}{12}
\ee
from \eq{C7tau1} that gives $c=1$ in both cases. We thus may expect that observables are universal.

Equation~\rf{C7a} is to be compared with an analogous formula for the free massive field $Y$
with the regularized propagator
\be
\LA Y(z) Y(0) \RA = 8\pi b_0^2 G_{M,a}(z),\qquad
G_{M,a}(z)=\frac1{2\pi }K_0\big(M\sqrt{z \bz+a^2}\big).
\label{YproGMa}
\ee
and the regularized delta-function
\be
\delta_{M,a} ^{(2)}(z) = (-4 \p\bp+M^2)\frac1{2\pi}K_0\big(M\sqrt{z \bz+a^2}\big)=
\frac{a^2 M^2K_2\big(M\sqrt{z \bz+a^2}\big)}{{2\pi}(z\bz+a^2)}.
\ee
In the limit $a M\to0$, which is justified  by small $a$ (and/or small $M$), we find
\be
\int \d^2 z\, \xi(z) \LA \p^n Y(z) Y(0) \RA \delta ^{(2)}(z)=(-1)^{n} {b_0^2} H_n\xi^{(n)}(0) 
%%\frac{1}{2^{n-1}n} \xi^{(n)}(0)
%\non H^{\rm (reg)}_1= 1,\qquad H^{\rm (reg)}_2= \frac{2}3,\qquad 
%H^{\rm (reg)}_3= \frac 12,\qquad H_n^{\rm (reg)}= \frac 2{n+1}
\label{dnYY}
\ee
reproducing the numbers in \eq{C7}. 
This shows the vanishing of the quantum correction to the total central charge 
of $\vp$ plus the requlator in the
quadratic case if the regulators are not  path-integrated out.
%%Equation~\rf{dnYY} is obtained 

One more sequence of the numbers emerges for the regularization by higher-derivatives like in \eq{inva}.
We then have 
%%\be
%%\LA \vp(-p) \vp(p) \RA = \frac{8\pi b_0^2}{p^2 +\eps p^4}.
%%\label{vppro}
%%\ee
for the  propagator of the  higher-derivative massless field 
%%in momentum space or
\be
\LA \vp(z) \vp(0) \RA = 8\pi b_0^2 G_{\eps}(z), \qquad
G_{\eps}(z)=-\frac1{2\pi}\left[K_0 \Big( \sqrt{\frac{z\bz}\eps} \Big)+ 
\log \big(\sqrt{z\bz}\mu \big)
\right]
\label{vpprox}
\ee
in coordinate space.  We also introduce
\be
\delta^{(2)}_{\eps}(z)=-4\p \bp  G_{\eps}(z)=\frac1{2\pi\eps} K_0 \left(\sqrt{\frac{z\bz}{\eps}}\right)
\stackrel{\eps\to0}\to
\delta^{(2)}(z)
\ee
to obtain
\bea
8\pi \int \d^2 z\, \xi(z) \p^{n} G_{\eps}(z) \delta^{(2)}_{\eps}(z)=
(-1)^{n} {b_0^2} H_n\xi^{(n)}(0)
\label{Geps1}
\eea
which is the same as for the regularization \rf{simplest} and
\bea
8\pi \int \d^2 z\, \xi(z)  (-4\eps\p^{n+1}\bp ) G_{\eps}(z) \delta^{(2)}_{\eps}(z)=
(-1)^{n} {b_0^2} J_n\xi^{(n)}(0),\non
J_1= 1,\qquad J_2= \frac{2}3,\qquad J_3= \frac 12,\qquad J_n = \frac 2{n+1}.
\label{Geps2}
\eea

%%\subsection{Higher derivative action}

For the  higher-derivative action~\rf{inva} %%as such 
the (unregularized)  propagator is given by 
 \rf{vpprox} but now we consider $\eps$ as a finite parameter. To be more precise, 
the consideration in the previous paragraph is associated with the usual BPZ case $\sqrt{\eps}\ll z$ 
when $\eps$ is a regularizing parameter, while
now we introduce a UV cutoff $a$ and consider $a\ll z \ll \sqrt{\eps}$. Speaking another way, we now
consider the limit $\eps\to\infty$ which is described for $G=0$ by a free conformal theory with two scalar fields~\cite{KN93}.

Introducing the UV cutoff $a$ as  before
\be
\LA \vp(z) \vp(0) \RA = 8\pi b_0^2 G_{\eps,a}(z), \qquad
G_{\eps,a}(z)=-\frac1{2\pi}\left[K_0 \Big( \sqrt{\frac{z\bz+a^2}\eps} \Big)+ 
\log\big(\sqrt{z\bz+a^2}\mu \big)
\right]
\ee
and also introducing
\be
\delta^{(2)}_{\eps,a}(z)=\left(-4\p \bp +16\eps \p^2\bp^2 \right) G_{\eps,a}(z)\stackrel{a\to0}\to
\delta^{(2)}(z),
\ee
we find after a little computation 
\bea
8\pi \int \d^2 z\, \xi(z) \p^n  G_{\eps,a}(z)\delta^{(2)}_{\eps,a}(z)\stackrel{a\to0}\to
0
\label{Geps0}
\eea
and
\bea
8\pi \int \d^2 z\, \xi(z)  (-4\eps\p^{n+1}\bp ) G_{\eps,a}(z)\delta^{(2)}_{\eps,a}(z)\stackrel{a\to0}\to
(-1)^{n} {b_0^2} Q_n\xi^{(n)}(0),\non
Q_1=1,\qquad Q_2=\frac3{10},\qquad Q_3=\frac2{15},\qquad Q_n=\frac{12}{n(n+2)(n+3)}.
\label{Gepsa}
\eea
This set of numbers differs from those in \eq{Geps2} except for $n=1$ which was the case
in \eq{44}.

The most general case is the higher-derivative massive field, whose  (unregularized) propagator reads
\be
\LA Y(-p) Y(p) \RA = \frac{8\pi b_0^2}{p^2+M^2 +\eps p^4}.
%%G_{1/M}(z),\qquad G_{1/M}(z)=\frac{1}{2\pi}K_0(M\sqrt{z \bz+a^2})
\label{Ypro}
\ee
in momentum space or, introducing the regularization $a$ as  before,
\bea
\LA Y(z) Y(0) \RA &= &8\pi b_0^2 G_{\eps,M,a}(z), \non
G_{\eps,M,a}(z)&=&\frac1{2\pi\eps(M_+-M_-)}\left[K_0 \big(  M_-\sqrt{z\bz+a^2} \big)-K_0 \big(  M_+\sqrt{z\bz+a^2} \big)
\right],\non
M_\pm&=&\frac{1\pm\sqrt{1-4\eps M^2}}{2\eps}
\eea
in coordinate space. The associated set of numbers coincides for $a\to0$ with the one in Eqs.~\rf{Geps0}, \rf{Gepsa}.
This again provides the cancellation of the (finite part of the) quantum correction to the central
charge coming from $\vp$ and the regulators if the regulators are not integrated out. 
The sum of two logarithmically divergent parts is then regularized as $\log (M^2\eps)$.

\end{document}